\documentclass[twocolumn,showpacs,floatfix]{revtex4}
\usepackage{latexsym}
\usepackage{graphics}
\usepackage{epsf}
\usepackage{epsfig}
\usepackage{graphicx}
\usepackage{dcolumn}
\usepackage{bm}
\usepackage{latexsym}
\usepackage{booktabs}
\usepackage{amssymb}
\usepackage[english]{babel}
\usepackage{hyperref}

\begin{document}

\noindent

\title{Reentrant Synchronization and Pattern Formation in Pacemaker-Entrained Kuramoto Oscillators}

\author{Filippo Radicchi\footnote{f.radicchi@iu-bremen.de} and Hildegard Meyer-Ortmanns\footnote{h.ortmanns@iu-bremen.de}}
\affiliation{School of Engineering and Science,
International University Bremen,\\
P.O.Box 750561, D-28725 Bremen, Germany}


\begin{abstract}
  We study phase entrainment of Kuramoto oscillators under
  different conditions on the interaction range and the natural
  frequencies. In the first part the oscillators are entrained by
  a pacemaker acting like an impurity or a defect.
  We analytically derive the entrainment frequency
  for arbitrary interaction range and the entrainment threshold for
  all-to-all couplings. For intermediate couplings our numerical results show
  a reentrance of the synchronization transition as a function of the
  coupling range. The origin of this reentrance can be traced
  back to the normalization of the
  coupling strength. In the second part we
  consider a system of oscillators with an initial gradient in
  their natural frequencies, extended over a one-dimensional chain
  or a two-dimensional lattice. Here it is
  the oscillator with the highest natural frequency that
  becomes the pacemaker of the ensemble, sending out circular
  waves in oscillator-phase space. No asymmetric coupling between
  the oscillators is needed for this dynamical induction of the
  pacemaker property nor need it be distinguished by a gap in the
  natural frequency.

\end{abstract}


\maketitle

\section{Introduction}
Synchronization in the sense of coordinated behavior in time is
essential for any efficient organization of systems, natural as
well as artificial ones. In artificial systems like factories the
sequence of production processes should be synchronized in a way
that time-and space consuming storage of input or intermediate
products is avoided. The same applies to natural systems like
networks of cells which obviously perform very efficiently in
fulfilling a variety of functions and tasks. In a more special
sense, the synchronized behavior refers to oscillators with almost
identical individual units, in particular to phase oscillators
with continuous interactions as described by the Kuramoto model
\cite{kurabook}. These sets of limit-cycle oscillators describe
synchronization phenomena in a wide range of applications
\cite{kuraothers}. One of these applications is pattern formation
in chemical oscillatory systems \cite{kuranakao}, described by
reaction-diffusion systems. A well-known example for such a
chemical system is the Belousov-Zabotinsky reaction, a mixture of
bromate, bromomalonic acid and ferroin, which periodically changes
its color corresponding to oscillating concentrations of the red,
reduced state and the blue, oxidized state. In these systems,
expanding target-like waves and rotating spiral waves are typical
patterns \cite{zaikin}. As Kuramoto showed (as e.g. in
\cite{kurabook}), these dynamical systems can be well approximated
by phase oscillators if the interaction is weak. For
reaction-diffusion equations with a nonlinear interaction term he
predicted circular waves with strong similarities to the
experimental observation of target patterns. In these systems,
defects or impurities seem to play the role of pacemakers, driving
the system into a synchronized state. Therefore he treated
pacemakers as local "defect" terms leading to heterogeneities in
the reaction-diffusion equations.

In this paper we consider two types of pacemakers. In the first
part they are introduced as "defects" in the sense that they have
a different natural frequency from the rest of the system, whose
oscillators have either exactly the same frequency or small random
fluctuations about some common average by assumption, where in the
second case the pacemaker's frequency is clearly different from
the values of a typical fluctuation. Actually, we will not
consider the latter case, since small fluctuations do not lead to
any qualitative change. The ad-hoc distinction of the pacemaker
may reflect natural and artificial systems with built-in
impurities like those in the Belousov-Zhabotinsky system. We have
recently studied these systems for nearest neighbor interactions
\cite{hmo1} and regular networks in $d$ dimensions for periodic
and open boundary conditions. For such systems we have shown that
it is the mean distance of all other nodes from the pacemaker (the
so-called depth of the network) that determines its
synchronizability. Periodic boundary conditions facilitate
synchronization as compared to open boundary conditions. Here we
extend these results to larger interaction range. The effect of an
extended interaction range on oscillatory systems without
pacemaker were considered in \cite{rogers}. We analytically derive
the entrainment frequency for arbitrary interaction range between
next-neighbor and all-to all interactions. We derive the
entrainment window for all- to-all couplings analytically, for
intermediate couplings numerically. The entrainment window depends
non-monotonically on the interaction range, so that the
synchronization transition is reentrant. In our system the
reentrance can be explained in terms of the weight of the
interaction term and the normalization of the coupling strength.

In the second part we consider a system of oscillators without a
"defect", all oscillators on the same footing up to the difference
in their natural frequencies. An obvious choice for the natural
frequencies would be a Gaussian distribution or another random
distribution to describe fluctuations in natural frequencies in
otherwise homogeneous systems, without impurities. Such systems
have been studied by Blasius and T\"onjes \cite{blasius} for a
Gaussian distribution. The authors showed that in this case an
asymmetric interaction term is needed to have synchronization,
this time driven by a dynamically established pacemaker, acting as
a source of concentric waves. This result shows that conditions
exist under which a system establishes its own pacemaker in a
"self-organized" way. In contrast to the random distribution of
natural frequencies we are interested in a (deterministic)
gradient distribution, in which the natural frequencies linearly
decrease over a certain region in space. Such a deterministic
gradient alone is probably not realistic for natural systems, but
considered as a subsystem of a larger set of oscillators with
natural frequency fluctuations it may be realized over a certain
region in space. Here we suppress the fluctuations and focus on
the effect of the gradient alone. As it turns out, the asymmetry
in the natural frequency distribution is sufficient for creating a
pacemaker as center of circular waves in oscillator-phase space,
without the need for an asymmetric term in the interaction. The
oscillator with the highest natural frequency becomes the source
of circular waves in the synchronized system. We call it
dynamically induced as its only inherent difference is its local
maximum at the boundary in an otherwise "smooth" frequency
distribution. The type of pattern created by the pacemaker has the
familiar form of circular waves. Beyond a critical slope in the
natural frequencies, full synchronization is lost. It is first
replaced by partial synchronization patterns with bifurcation in
the frequencies of synchronized clusters, before it gets
completely lost for too steep slopes.

The outline of the paper is as follows. In the first part we treat
the pacemaker as defect and derive the common entrainment
frequency for arbitrary interaction range and topology (section
\ref{sec.I}). Next we determine the entrainment window,
analytically for all-to all coupling and numerically for
intermediate interaction range (section \ref{sec.I1}). Here we see
the reentrance of the transition as a function of the interaction
range (section \ref{sec.I2}). In the second part (section
\ref{sec.II}) we consider dynamically induced pacemakers, first
without asymmetric interaction term, for which we analytically
derive the synchronization transition as a function of the
gradient in the frequencies (sections \ref{sec.II1}). In section
\ref{sec.II2} we add an asymmetric term, so that the pattern
formation is no longer surprising due to the results of
\cite{blasius}, but the intermediate patterns of partial
synchronization are different. In section \ref{sec.IV} we
summarize our results and conclusions.
\section{Pacemaker as defect in the system}
\label{sec.I}

The system is defined on a network, regular or small-world like.
To each node $i$, $i=0,\ldots, N$, we assign a limit-cycle
oscillator, characterized merely by its phase $\varphi_i$, which
follows the dynamics
\begin{equation}
\dot{\varphi}_i = \omega + \delta_{i,s} \Delta \omega+ \frac{K}{k_i} \sum_{j\not=i} A_{j,i}
\sin{\left(\varphi_j-\varphi_i\right)}\;, \label{def:model}
\end{equation}
with the following notations. The frequency $\omega$ denotes the
natural frequencies of the system. In this system we treat the
pacemaker as a defect. It is labelled by $s$ and has a natural
frequency that differs by $\Delta\omega$ from the frequency of the
other oscillators having all the same frequency $\omega$.
$\delta_{i,j}$ denotes the Kronecker delta. The constant $K>0$
parameterizes the coupling strength. We consider regular networks
and choose
\[A_{j,i}=r^{-\alpha}_{j,i}\;,
\]
$r_{j,i}$ the distance between nodes $j$ and $i$, that is
\begin{equation}
r_{j,i}=\min\left( \left[ \; \vert j-i\vert \; , \; (N+1)-\vert
j-i\vert \; \right]\right) \label{def:distance}
\end{equation}
on a one-dimensional lattice with periodic boundary conditions. In
two or higher dimensions it is the shortest distance in lattice
links. The parameter $0\leq\alpha\leq\infty$ tunes the interaction
range. Alternatively we consider $A_{j,i}$ as the adjacency matrix
on a small-world topology: $A_{j,i}=1$ if the nodes $j$
and $i$ are connected and $A_{j,i}=0$ otherwise). Moreover,
$k_i=\sum_{j}A_{j,i}$ is the degree of the $i$-th node, it gives
the total number of connections of this node in the network. This
system was considered before in \cite{hmo1} for nearest-neighbor
interactions ($\alpha \to \infty$) on a $d$-dimensional hypercubic
lattice and on a Cayley tree. Here we extend the results to
long-range interactions via $0\leq\alpha<\infty$.

\subsection*{Entrainment frequency and entrainment window} \label{sec.I1}
In the appendix of \cite{hmo1} we derived the common entrainment
frequency $\Omega$ in the phase-locked state, for which
$\dot{\varphi}_i \equiv \Omega$ for all $i$, to be given as
\begin{equation}
\Omega\;=\;\Delta\omega \frac{k_s}{ \sum_{i} k_i}
\end{equation}
in the rotated frame, in which the natural frequency $\omega$ is
zero. In particular such a result is obtained directly from system
(\ref{def:model}), applying the transformation to the rotating
frame $\varphi_i \to \varphi_i+\omega t$ $\forall i$, multiplying
the $i$-th equation by $k_i$, summing over all equations and then
using the fact that $\sum_i \sum_j A_{j,i} \sin{(
\varphi_j-\varphi_i) }=0$, because the symmetry of the adjacency
matrix $A$ and the antisymmetry of the $\sin$-function. Whenever
$k_i\equiv k$ is independent of $i$ as it happens for periodic
boundary conditions and homogeneous degree distributions, in
particular for $\alpha=0$ (all-to-all coupling), the common
frequency (in the non-rotated frame) is given by
\[
\Omega  \;=\;\frac{\Delta\omega}{N+1} \;+\;\omega\;,
\]
independent on $\alpha$. In the limit $N\rightarrow\infty$  the
system synchronizes at $\omega$.
\\
Analytical results for the entrainment window were derived for
open and periodic boundary conditions and $\alpha\rightarrow
\infty$ in \cite{hmo1}. For $\alpha=0$, the result is the same as
for $\alpha\rightarrow\infty$, periodic boundary conditions and
$d\geq 1$
\begin{equation}
\left\vert\frac{\Delta\omega}{K}\right\vert_c\;=\;\frac{1}{N} +1
\;\;\;. \label{eq:critic}
\end{equation}
This is seen as follows. Since for all-to-all couplings the
mean-field approximation becomes exact, it is natural to rewrite
the dynamical equations (\ref{def:model}) in terms of the order
parameter $Re^{i\psi}$ defined according to
\[R\;e^{i\psi}\;=\;\frac{1}{N}\sum_{j\not=s}e^{i\varphi_j}\;.
\]
This quantity differs from the usual order parameter just by the
fact that the sum runs over $j\not=s$. Eq.s(\ref{def:model}) then take
the form
\begin{equation}
\begin{array}{ll}
\Omega =\Delta\omega\;+\;K\;R\;\sin(\psi-\varphi_s) & \textrm{ ,
for } i=s
 \\
\Omega=\frac{K}{N}\sin(\varphi_s-\varphi_i)\;+\;K\;R\;\sin(\psi-\varphi_i)
& \textrm{ , }\forall \; i\not= s\
\end{array}
\; , \label{eq.meanfield}
\end{equation}
Using $\varphi_i=\psi$ for all $i\not= s$ as one of the possible
solutions and the fact that the $\sin$-function is bounded, these
equations imply the entrainment window as given by
(\ref{eq:critic}) with $N$ the total number of nodes minus $1$. As
we have shown in \cite{hmo1}, the same dependence in terms of $N$
holds for a $d$-dimensional hypercubic lattice with $N^\prime$ nodes
per side and nearest neighbor couplings, for which
$N=(N^\prime+1)^d-1$, so that the entrainment window for nearest
neighbor interactions and on a regular lattice decays
exponentially with the depth of the network in the limit of
infinite dimensions $d$ ; for random and small-world networks the
same type of decay was derived in \cite{kori}.

The intermediate interaction range $0<\alpha<\infty$ cannot be
treated analytically. Here the numerical integration of the
equations (\ref{def:model}) (via a fourth-order Runge-Kutta
algorithm with time step $dt=0.05$ and homogeneous initial
conditions) shows that the entrainment window is even smaller than
in the limiting cases $\alpha=0$ or $\alpha=\infty$ for otherwise
unchanged parameters (that is fixed size $N$, dimension $d$,
coupling $K$). For these cases formula (\ref{eq:critic}) gives an
upper bound on the entrainment window, as one can easily see from
Eq.\ref{def:model} for $i=s$ because $\sum_{j \neq s}
\sin{(\varphi_j - \varphi_s)}/r_{j,s}^{\alpha}\leq \sum_{j \neq s}
1/r_{j,s}^{\alpha}= k_s$ ,  $\forall \alpha$. Actually the
entrainment window becomes smaller for intermediate interaction
range, a feature leading to the reentrance of the synchronization
transition as function of $\alpha$, as we shall see in the figures
\ref{fig:reentalpha} and \ref{fig:reentkandp} below.

\subsubsection*{Reentrance of the phase transition}\label{sec.I2}
\begin{figure}[ht]
\includegraphics*[width=0.47\textwidth]{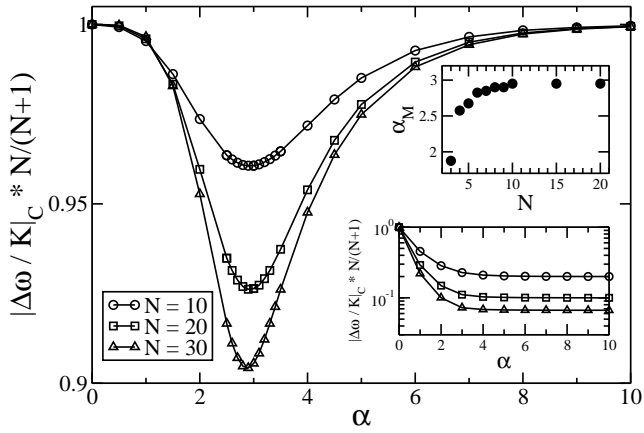}
\caption{Normalized entrainment window as function of the
interaction range parameterized via $\alpha$, for a
one-dimensional lattice with $N+1$ sites, periodic boundary
conditions and for interaction strength between the oscillators
$i$ and $j$ suppressed by the power $\alpha$ of their distance
$r_{j,i}$. The main plot indicates the reentrance of the
synchronization transition. The upper inset shows the size
dependence of $\alpha_m$, for which the entrainment window becomes
minimal, it saturates at $\alpha\sim 3$. The lower inset shows the
monotonic dependence on $\alpha$ for constant coupling strength
$K/N$. The figure was obtained for homogeneous initial conditions.
The qualitative features remain the same for random initial
conditions.} \label{fig:reentalpha}
\end{figure}

\begin{figure}[ht]
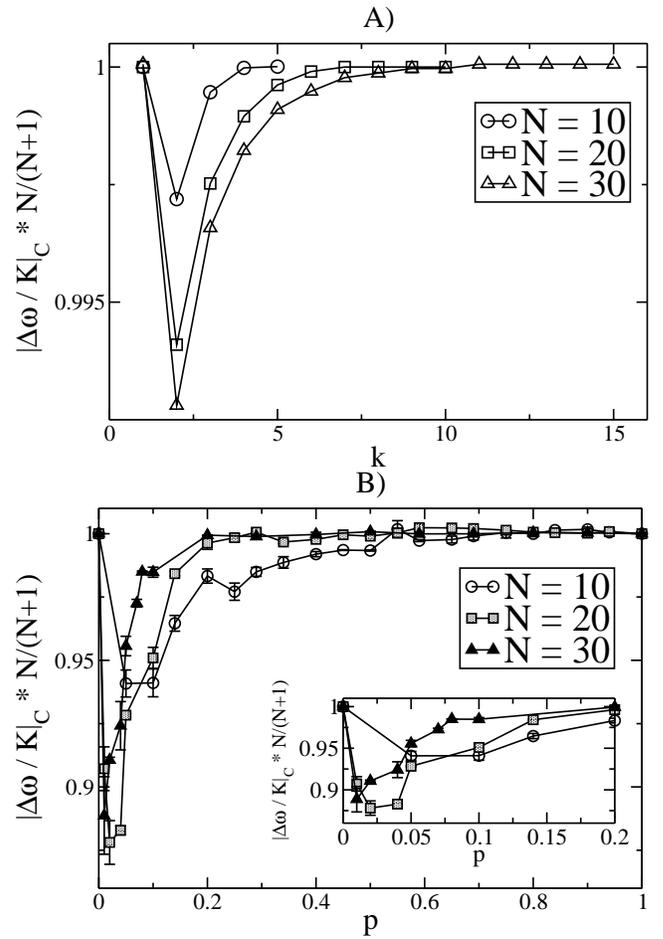

\includegraphics*[width=0.47\textwidth]{reentkandp1}
\includegraphics*[width=0.47\textwidth]{reentkandp2}
\caption{Normalized entrainment window  as function of the number
of neighbors $k$ (A) and of the probability $p$ for adding
shortcuts in a small-world topology (B) for a one-dimensional
lattice with periodic boundary conditions and $N+1$ sites. Average
values in (B) are taken over $50$ realizations, error bars
correspond to the standard deviation. Again the non-monotonic
behavior indicates reentrance of the synchronization transition.}
\label{fig:reentkandp}
\end{figure}

Reentrance phenomena are observed in a variety of phase
transitions, ranging from superconductor-insulator transitions as
function of temperature \cite{vanotterlo} and noise-induced
transitions \cite{castro1995} to chaotic coupled-map lattices as
function of the coupling \cite{anteneodo}. In \cite{kampf} a
reentrance phenomenon is discussed as an artifact of the
approximation. Reentrance of phase transitions is challenging as
long as it appears counterintuitive. For example, it is
counterintuitive when synchronization is first facilitated for
increasing coupling and later gets blocked when the coupling
exceeds a certain threshold, so that synchronization depends
non-monotonically on the coupling. As we see in Fig.
\ref{fig:reentalpha} for a one-dimensional system with periodic
boundary conditions and for various system sizes $N$, the
entrainment window $\vert\Delta\omega/K\vert_c$ depends
non-monotonically on the interaction range, parameterized by
$\alpha$. This dependence is easily explained by looking at the
total weight coming from the sum of interactions. This
contribution depends non-monotonically on $\alpha$, as
$\sum_j\sin(\phi_j-\phi_i)/r_{j,i}^\alpha$ decreases for
increasing $\alpha$, while the coupling strength increases under
the same change of $\alpha$, so that there is a competition
between the two factors leading to a minimum of the weight for a
certain value of $\alpha$. In contrast, if we scale the coupling
strength by a constant factor $K/N$, independently of the degree
of the node $i$, the overall weight of the sum decreases for
increasing $\alpha$ and along with this, the drive to the
synchronized state. As it is shown in the inset of Fig.
\ref{fig:reentalpha}, the threshold monotonically decreases with
$\alpha$ if the coupling strength is normalized by $1/N$
independently of the actual degree, as expected from our arguments
above. The second inset shows the $N$-dependence of the value of
$\alpha$ for which synchronization is most difficult to achieve,
it converges to $\alpha=3$ for large $N$. Similar non-monotonic
behavior is observed if the interaction range is tuned via the
number $k$ of nearest neighbors on a ring rather than via
$\alpha$, see Fig. \ref{fig:reentkandp}A and for a small-world
topology \cite{watts}, see Fig. \ref{fig:reentkandp}B. Here the
small-world is constructed by adding random shortcuts with
probability $p$ from a regular ring with nearest neighbor
interactions ($p=0$) to an all-to-all topology ($p=1$)
\cite{newman}.


\section{Dynamically induced pacemakers}\label{sec.II}

In view of dynamically induced pacemakers we study the dynamical
system
\begin{equation}
\dot{\varphi}_i=\omega_i+\frac{K}{k_i} \sum_{j} A_{j,i}
\Gamma(\varphi_j-\varphi_i)\;\;\; , \label{eq1}
\end{equation}
without a pacemaker. The interaction term $\Gamma$ is given by
\begin{equation}
\Gamma{\left(\varphi_j-\varphi_i\right)}\;=\;\sin
\left(\varphi_j-\varphi_i\right)\;+\;
\gamma\left[1\;-\;\cos\left(\varphi_j-\varphi_i\right)\right]\;,
\label{def:gamma}
\end{equation}
 where $\gamma = 0$ corresponds to antisymmetric interaction,
$\gamma>0$ to asymmetric interaction. In the following we consider
the cases $\gamma=0$ and $\gamma=2$. The value $\gamma=2$ is
chosen for convenience in order of comparing our results with
those of Blasius and T\"onjes \cite{blasius}, but different values
of $\gamma>0$ lead to qualitatively the same features.
\\
We focus on lattices with periodic boundary conditions in one and
two dimensions. Moreover we consider only nearest neighbor
interactions, so that in  $d=1$ we have $k_i=2$ $\forall i$ and
 $k_i=4$ $\forall i$ in $d=2$. The natural frequencies $\omega_i$
are chosen from a gradient distribution\begin{equation} \omega_i =
\Delta\omega - r_{i,s} \;
\frac{\Delta\omega}{\max_i{r_{i,s}}}\;\;\; , \label{eq2}
\end{equation}
where $r_{i,j}$ is the network distance between the oscillators
$i$ and $j$, given by (\ref{def:distance}) in one dimension and by
the shortest path along the edges of the square lattice in two
dimensions. The oscillator $s$ is the oscillator with the highest
natural frequency. For a one-dimensional lattice with periodic
boundary conditions (ring) and  $N+1$ oscillators, choosing $s=0$
for simplicity, $\max_i{r_{i,0}}= N'$, with $N'=N/2$ for $N$ even
and $N'=(N+1)/2$ for $N$ odd.  In the following we choose $N$ as
even; the case of $N$ odd is very similar.

\subsection{Pattern formation for $\gamma=0$ }\label{sec.II1}
\subsubsection{Entrainment frequency}\label{sec.II11}
Next we derive the common frequency in case of
phase-locking $\dot{\varphi}_i\equiv \Omega$ , $\forall i$.
Summing the equations of (\ref{eq1}) and using the antisymmetry of
$\Gamma$ for $\gamma = 0$, we obtain
\[
\begin{array}{l}
(N+1) \Omega = \sum_{i=0}^{N} \omega_i = \Delta\omega + 2
\Delta\omega \; \sum_{i=1}^{N/2} \left(1-2i/N\right)=
\\
= \Delta\omega \left[ 1 +N - 4/N \; \frac{N/2(N/2+1)}{2}\right] =
\Delta\omega \frac{N}{2}\;,
\end{array}
\]
leading to
\begin{equation}
\Omega = \Delta\omega \frac{N}{2(N+1)}\;, \label{eq:common}
\end{equation}
so that  $\Omega\;\rightarrow\;\Delta\omega/2$ for
$N\rightarrow\infty$.

\subsubsection{Entrainment window}\label{sec.II12}
It is convenient to rewrite the differential equations (\ref{eq1})
in terms of phase lags $\theta_i=\varphi_{i}-\varphi_{i-1}$, so
that the equation for the oscillator at position $0$ reads
\[
\Omega = \omega_0 + K \sin{(\theta_1)} \; \Rightarrow \;
\sin{(\theta_1)} = \frac{\Omega-\omega_0}{K}\; .
\]

\begin{figure}
\includegraphics*[width=0.47\textwidth]{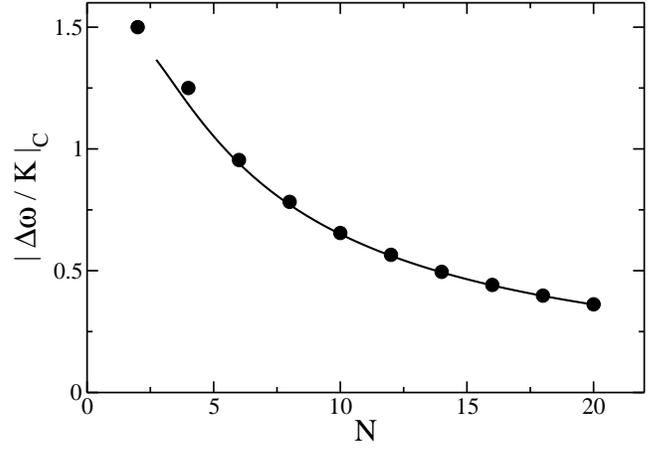}
\caption{Critical threshold for a one-dimensional lattice as
function of its linear size $N$, full line corresponds to the
analytic prediction, full dots to numerical results for $N\geq2$,
the symmetry breaking parameter is chosen as $\gamma=0$.}
\label{fig.theonum}
\end{figure}

For the other oscillators we obtain
\[
\begin{array}{l}
\Omega =  \omega_1  + K/2 \left[ \sin{(\theta_2)} -
\sin{(\theta_1)} \right]
\\
\Rightarrow \;
\sin{(\theta_2)} = 2\frac{\Omega-\omega_1}{K} + \sin{(\theta_1)}
\end{array}
\;\;\; ,
\]

\[
\begin{array}{l}
\Omega =  \omega_2  + K/2 \left[ \sin{(\theta_3)} -
\sin{(\theta_2)} \right]
\\
\Rightarrow \;
\sin{(\theta_3)} = 2\frac{\Omega-\omega_2}{K} + \sin{(\theta_2)}
\end{array}
\;\;\; ,
\]

so that in general
\[
\begin{array}{l}
\Omega =  \omega_{i-1}  + K/2 \left[ \sin{(\theta_{i})} -
\sin{(\theta_{i-1})} \right]
\\
\Rightarrow \; \sin{(\theta_{i})} = 2
\frac{\Omega-\omega_{i-1}}{K} +  \sin{(\theta_{i-1})}
\end{array}
\]
or
\[
\\
\sin{(\theta_{i})} = \sum_{j=1}^{i-1} \; 2
\frac{\Omega-\omega_j}{K} +  \sin{(\theta_1)}\;.
\]
The former sum can be performed as
\[
\begin{array}{l}
\sum_{j=1}^{i-1} \; 2 \frac{\Omega-\omega_j}{K} = \frac{2}{K} \;
\sum_{j=1}^{i-1} \left[\Omega - \Delta\omega
\left(1-2j/N\right)\right]=
\\
 = \frac{2}{K} \left( \Omega  - \Delta\omega\right) (i-1) + \Delta\omega
 \frac{2}{K} \frac{i(i-1)}{N}\;,
\end{array}
\]
from which
\[
\sin{(\theta_{i})} = \frac{\Omega  - \Delta\omega}{K}  \; (2i-1) +
\frac{\Delta\omega}{K} \frac{2i^2-2i}{N}\;,
\]
or, using Eq.(\ref{eq:common}),
\begin{equation}
\begin{array}{l}
\sin{(\theta_{i})} = \frac{\Delta\omega}{K} (2i-1)
\left[\frac{N}{2(N+1)} -1\right] +  \frac{\Delta\omega}{K}
\frac{2i(i-1)}{N} = \\
=  \frac{\Delta\omega}{K} \; \left[ (1-2i) \frac{N+2}{2(N+1)} +
\frac{2i(i-1)}{N}\right] \;.
\end{array} \label{eq:critic1}
\end{equation}
For given $N$, we first determine  the position $i_m$ at which the
right-hand-side of Eq.(\ref{eq:critic1}) takes its minimum
(maximum) value as a function of $i$. This value is constrained by
the left-hand-side of (\ref{eq:critic1}) to be out of $[-1,1]$.
For $i$ treated as a continuous variable the derivative of
Eq.(\ref{eq:critic1}) with respect to $i$ yields
\begin{equation}
i_m= \frac{N(N+2)}{4(N+1)}+\frac{1}{2} \;.\label{eq:4a}
\end{equation}
On the other hand
\begin{equation}
i_m \geq 1 \; \Leftrightarrow  N \geq \sqrt{2} \;,
\end{equation}
and \begin{equation}
 i_m \leq \frac{N}{2} \; \Leftrightarrow \;  N
\geq 1+\sqrt{3} \;,\label{eq.11a}
\end{equation}
so that for $N \gg 1$ , $1 \leq i_m \leq N/2$ is certainly
satisfied. Upon inserting $i_m$ from (\ref{eq:4a}) we have
\[
\begin{array}{l}
\left|\frac{\Delta\omega}{K}\right|_C = \left| (1-2i_m)
\frac{N+2}{2(N+1)} + \frac{2i_m(i_m-1)}{N}\right|^{-1}=
\\
=\left( \frac{N^2(N+2)^2+4(N+1)^2}{8N(N+1)^2} \right)^{-1}\;.
\end{array}
\]
The critical threshold is then given by
\begin{equation}
\left|\frac{\Delta\omega}{K}\right|_C =
\frac{8N(N+1)^2}{N^2(N+2)^2+4(N+1)^2}\;. \label{eq:critic_fin}
\end{equation}
This relation is plotted in Fig. \ref{fig.theonum}, where the full
line represents the theoretical prediction, while the full dots
correspond to the threshold values, numerically determined by
integrating the system of differential Eq.s (\ref{eq1}), starting
with $N=2$. (The case of $N=2$ violates the constraint
(\ref{eq.11a}) in agreement with the fact that for $N=2$ the
system degenerates to the former case of having a pacemaker as a
defect and two oscillators with natural frequencies
$\omega_1=\omega_2=0$.) Furthermore it should be noticed that
Eq.(\ref{eq:critic_fin}) scales as $1/N$ for large values of $N$,
the large-$N$ behavior is, however, not visible for the range of
$N$, ($N \leq 20$), plotted in Fig. \ref{fig.theonum}.

\begin{figure}
\includegraphics*[width=0.47\textwidth]{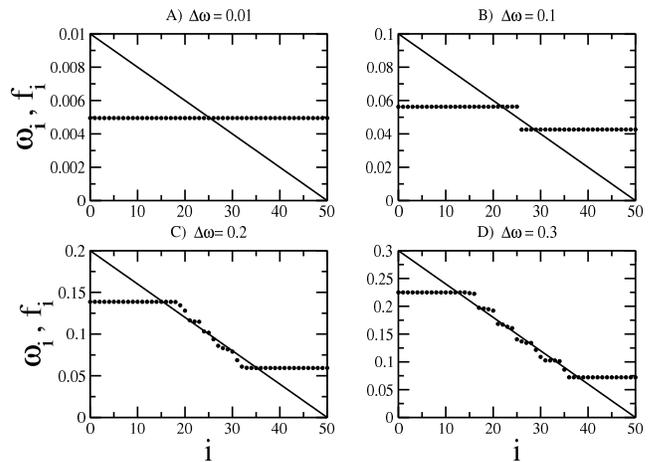}
\caption{Natural frequencies $\omega_i$ (full lines) and measured,
stationary frequencies $f_i$ (full dots) as function of the
position $i$ of the oscillators on a one-dimensional lattice, for
$N=100$, periodic boundary conditions, and four values of the
maximal natural frequency $\Delta\omega$. The oscillator with
maximal natural frequency $\Delta\omega$ is at position $s=0$.
Only half of the oscillators up to index $50$ are shown because of
the symmetric arrangement. The symmetry breaking parameter is
chosen as $\gamma = 0$.} \label{fig.former7}
\end{figure}

\begin{figure}
\includegraphics*[width=0.47\textwidth]{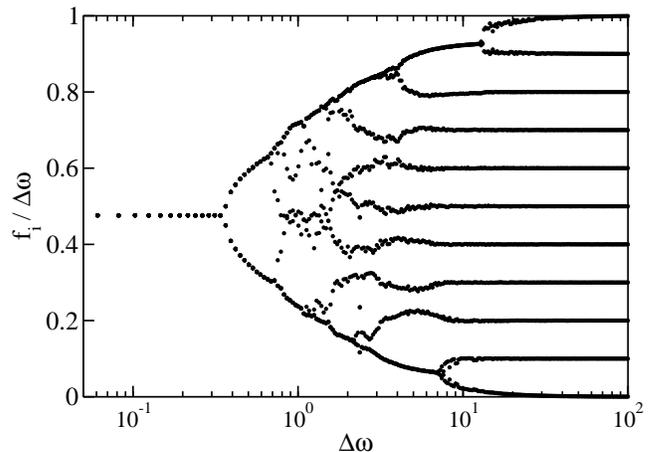}
\caption{Bifurcation in cluster-frequency space above a critical
slope in the natural frequencies, parameterized by $\Delta\omega$,
the maximal natural frequency. Numerically measured frequencies
$f_i$ (normalized such that $f_i/\Delta\omega$ $\in$ $[0,1]$) as
function of $\Delta\omega$ for a one-dimensional lattice with
$N=20$ and periodic boundary conditions. The symmetry breaking
parameter is chosen as $\gamma =0$.} \label{fig.bif}
\end{figure}

\begin{figure}
\includegraphics*[width=0.47\textwidth]{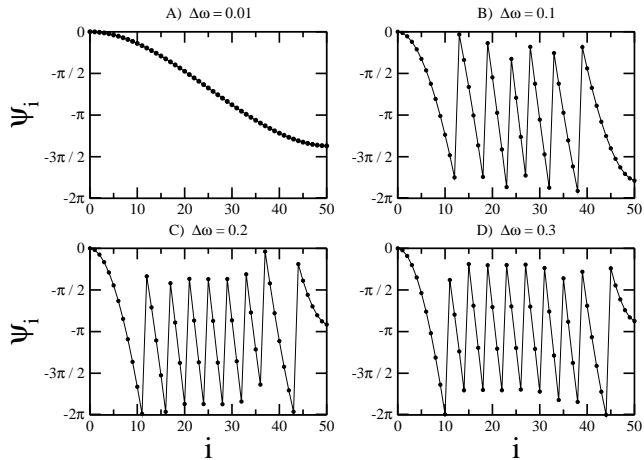}
\caption{Phase differences $\psi_i=\varphi_i-\varphi_0$ (from the
oscillator at $i=0$) as function of the oscillator index $i$ for a
one-dimensional lattice, with $N=100$ and periodic boundary
conditions, for different values of the maximal natural frequency
$\Delta\omega$ and $\gamma = 0$. The difference depends
nonlinearly on the distance from the oscillator at $i=0$. The
``steps'' in these plots are due to the projection of the
variables $\psi_i$ onto the interval $[0,2\pi)$.}
\label{fig.former8}
\end{figure}

For a ring with maximal distance $N^\prime = N/2$ and given $N$
and $K$, the slope $\Delta \omega/N^\prime$ in the natural
frequencies determines whether synchronization is possible or not.
The oscillators synchronize if the values of $N$ lead to a ratio
$\vert\Delta\omega/K\vert$ below the bound
$\vert\Delta\omega/K\vert_C$ of Eq.(\ref{eq:critic_fin}), named as
in section \ref{sec.I}, Eq.(\ref{eq:critic}), but with
$\Delta\omega$ now standing for the maximal difference between the
highest and the lowest natural frequencies. If for given $N$ the
slope, parameterized by $\Delta\omega$, exceeds this threshold, no
entrainment is possible. For $N\rightarrow\infty$ the allowed
slope goes to zero, the system does no longer synchronize. To
further characterize the synchronization patterns, we measure the
stationary frequency $f_i$ of the individual oscillators $i$,
defined according to
\begin{equation}
f_i = \lim_{t \to \infty}
\frac{\varphi_i(t+t_0)-\varphi_i(t_0)}{t-t_0} \label{eq3}\;,
\end{equation}
as well as its average $f\;=\;\frac{1}{N+1}\sum_{i=0}^N f_i$ and
variance
$\sigma\;=\;\left[\frac{1}{N+1}\sum_{i=0}^{N}\left(f_i-f\right)^2\right]^{1/2}$.
The average frequency and variance are introduced to characterize
the intermediate patterns of partial synchronization and to
distinguish it from the case $\gamma=2$ in section \ref{sec.II2}.
In all numerical calculations we set from now on $K=1$. Fig.
\ref{fig.former7} shows the natural frequencies $(\omega_i)$
(represented as full line) and the stationary $(f_i)$ frequencies
(shown as dots) for four slopes $\Delta \omega = 0.01, 0.1, 0.2,
0.3$, respectively, as function of the position on a linear chain
of $100$ oscillators, of which we show only one half due to the
symmetric arrangement. The case of $\Delta\omega = 0.01$
corresponds to full synchronization, while the other three figures
represent partial synchronization of two $(\Delta\omega=0.1)$ and
more clusters. It should be noticed that the two clusters for
$(\Delta\omega=0.1)$ would violate the condition
(\ref{eq:critic_fin}) if they were isolated clusters of $50$
oscillators each with a slope determined by $\Delta\omega = 0.1$,
but due to the nonlinear coupling to the oscillators of the second
cluster, the partial synchronization is a stationary pattern. Fig.
\ref{fig.bif} displays the frequencies $f_i$ as function of the
slope, again parameterized by $\Delta\omega$. Here we have chosen
$N=20$. We clearly see the bifurcation in frequency space starting
at the critical value of $\Delta\omega \simeq 0.36$ and ending
with complete desynchronization, in which the stationary
frequencies equal the natural frequencies. For frequency
synchronization the phases are locked, but their differences to
the phase of say the oscillator $s=0$
($\psi_i=\varphi_i-\varphi_0$) increases nonlinearly with their
distance from $s$, as seen in Fig. \ref{fig.former8}. Therefore
the distance between points of the same $\psi$ decreases with $i$.

\begin{figure}
\includegraphics*[angle=270,width=0.47\textwidth]{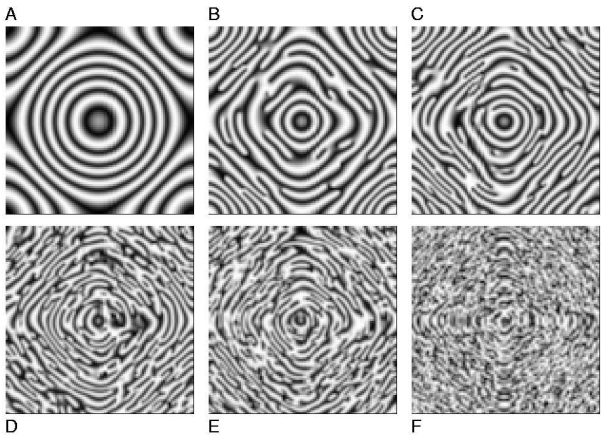}
\caption{Pattern formation for symmetry breaking parameter
$\gamma=0$ on a two-dimensional lattice with $N=100$ oscillators
per side. The oscillator $s$ with the maximal natural frequency is
placed at the center the lattice. We plot $\sin \psi_i$ for every
oscillator $i$. The colors vary between white, corresponding to
$-1$, and black, corresponding to $1$. The plots are realized
after $2 \cdot 10^4$ integration steps and for different values of
$\Delta\omega$ [therefore after different time in natural units of
$\Delta\omega$]: A) $0.1$ [$10^2$] , B) $0.5$ [$5 \cdot 10^2$] ,
C) $1.0$ [$10^3$] , D) $5.0$ [$5 \cdot 10^3$] , E) $10$ [$10^4$] ,
F) $50$ [$5\cdot 10^4$]. A cross section through the pattern of A)
would reveal a frequency distribution as in
Fig.\ref{fig.former7}B.} \label{fig.former9}
\end{figure}

\begin{figure}
\includegraphics*[angle=270,width=0.47\textwidth]{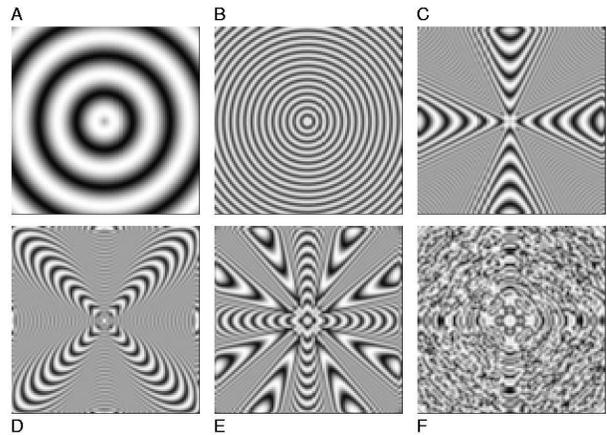}
\caption{Time evolution of the "disorder" in Fig.
\ref{fig.former9}F for symmetry breaking parameter $\gamma=0$. All
plots are realized for $\Delta\omega=50$, after a different number
of integration steps [time in natural units of $\Delta\omega$]: A)
$9$ [$22.5$] B) $41$ [$102.5$] , C) $150$ [$375$] , D) $393$
[$982.5$] , E) $490$ [$1225$], F) $1995$ [$4987.5$]. Only the
pattern of F) is stationary on a global scale, i.e., up to local
fluctuations in the phases.} \label{fig.former10}
\end{figure}

In two dimensions we simulated $100\times 100$ oscillators on a
square lattice with periodic boundary conditions and the
oscillator $s$ with the highest natural frequency placed at the
center of the square lattice. The natural frequencies of the
oscillators are still given by Eq.(\ref{eq2}). Their spatial
distribution looks like a square pyramid centered at the middle of
the square lattice. Fig. \ref{fig.former9} displays $\sin\psi_i$
on this grid. It exhibits stationary patterns after $2 \cdot 10^4$
steps of integration, for six choices of $\Delta\omega$, all above
the synchronization threshold. We see some remnants of
synchronization, most pronounced in Fig. \ref{fig.former9}A, in
which a circular wave is created at the center at $s$, and
coexists with waves absorbed by sinks at the four corners of the
lattice. The projection of Fig. \ref{fig.former9}A on one
dimension corresponds to Fig. \ref{fig.former7}B with a
bifurcation into two cluster-frequencies. Note that it is again
the oscillator with the highest natural frequency that becomes the
center of the outgoing wave, while the corners with the lowest
natural frequencies $(\omega_i=0)$ become sinks. Although Fig.
\ref{fig.former9}F shows almost no remnant of synchronization, it
is interesting to follow the time evolution towards this
"disordered" pattern via a number of snapshots, as displayed in
Fig. \ref{fig.former10} after
  $9$ [$22.5$] (A), $41$ [$102.5$] (B) , $150$ [$375$] (C) , $393$ [$982.5$] (D),
$490$ [$1225$] (E), $1995$ [$4987.5$] (F) integration steps $T$
[$T\;\Delta\omega$ time steps in units of natural
frequency($\Delta\omega)$ of Eq.\ref{eq2}]. While the pattern of
(A) would be stationary in case of full synchronization, here it
evolves after iterated reflections to that of (F). The pattern of
(F) is stationary in the sense that it stays "disordered", on a
"microscopic" scale it shows fluctuations in the phases.  The
evolution takes a number of time steps, since the interaction is
mediated only via nearest neighbors and not of the mean-field type
(all-to-all coupling).

\subsection{Pattern formation for $\gamma=2$}\label{sec.II2}
\begin{figure}
\includegraphics*[width=0.47\textwidth]{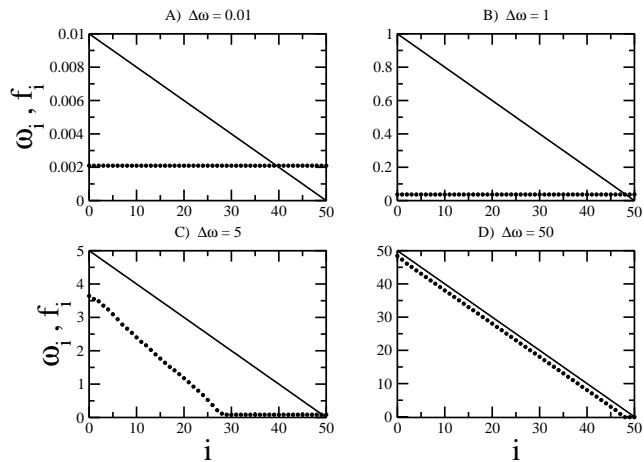}
\caption{Natural and measured frequencies for symmetry breaking
parameter $\gamma=2$. The frequency of the synchronized cluster
decreases and the cluster size shrinks, until all oscillators keep
their natural frequency in a completely desynchronized state.}
\label{fig.former3}
\end{figure}

\begin{figure}
\includegraphics*[width=0.47\textwidth]{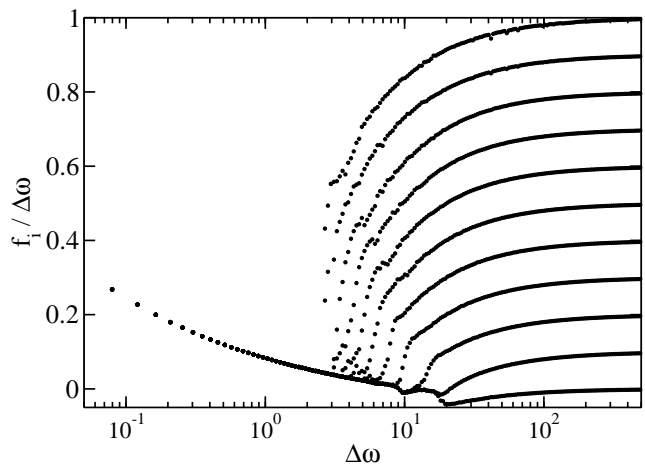}
\caption{Bifurcation in cluster-frequency for symmetry breaking parameter $\gamma=2$. Above the
synchronization threshold the size of the original synchronized
cluster shrinks with $\Delta\omega$. Oscillators outside this
cluster stay isolated in contrast to Fig. \ref{fig.bif}.}
\label{biforc2}
\end{figure}

\begin{figure}
\includegraphics*[angle=270,width=0.47\textwidth]{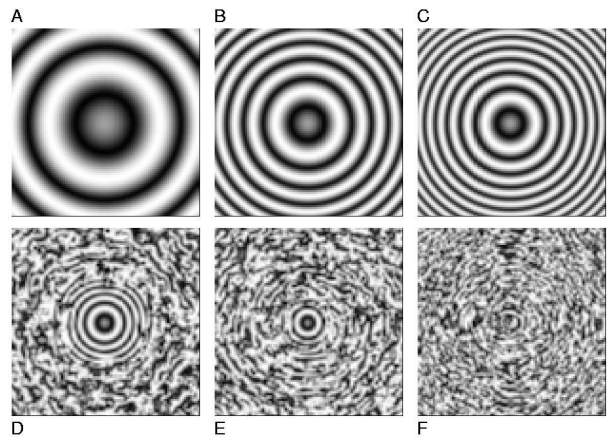}
\caption{Pattern formation for symmetry breaking parameter $\gamma=2$. We use the same
parameter choice as in Fig. \ref{fig.former9}. Now Fig.s A), B),
C) correspond to full synchronization, while one-dimensional cross
sections through D), E), and F) would reveal distributions of
$f_i$ as in Fig.\ref{fig.former3} C)-D).} \label{fig.former5}
\end{figure}

\begin{figure}
\includegraphics*[angle=270,width=0.47\textwidth]{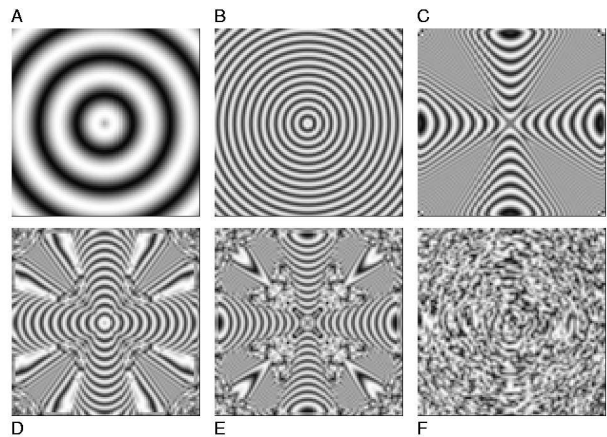}
\caption{Time evolution for symmetry breaking parameter
$\gamma=2$, otherwise same as in Fig. \ref{fig.former10}. This
time the plots are realized after $9$  [$22.5$] (A), $48$ [$120$]
(B), $160$ [$400$] (C), $277$ [$692.5$] (D), $386$ [$965$] (E),
and $2926$ [$7315$] (F) time steps [time in units of
$\Delta\omega$]. Only the disordered pattern of (F) is stationary
up to local fluctuations in the phases.} \label{fig.former6}
\end{figure}

Next let us consider system (\ref{eq1}) for $\gamma=2$, for which
we expect pattern formation from the results of \cite{blasius} due
to the broken antisymmetry of $\sin$ via a $\cos$-term in the
interaction. The main difference shows up in the partial
synchronization patterns above the synchronization threshold. As
it is evident from Fig.s \ref{fig.former3} and \ref{biforc2}, the
size of the one synchronized cluster, filling the whole lattice
below the threshold, shrinks with increasing $\Delta\omega$, since
the oscillators with the highest natural frequencies decouple from
the cluster, but the remaining oscillators do not organize in
other synchronized clusters, the cluster-frequency does no longer
bifurcate as before. Along with this, the two-dimensional
stationary patterns change to those of Fig. \ref{fig.former5},
whereas the evolution towards Fig. \ref{fig.former5}F is now
displayed in Fig.s \ref{fig.former6} with snapshots taken after
 $9$  [$22.5$] (A),
$48$ [$120$] (B), $160$ [$400$] (C), $277$ [$692.5$] (D), $386$
[$965$] (E), and $2926$ [$7315$] (F) integration steps
[$T\;\Delta\omega$ as above].

\begin{figure}
\includegraphics*[width=0.47\textwidth]{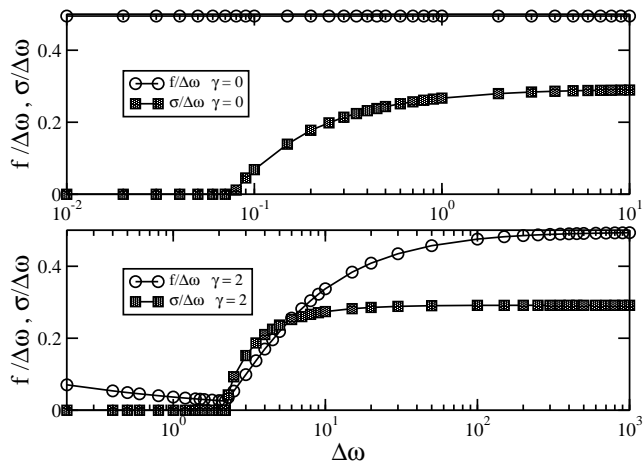}
\caption{Average frequency and variance as defined in the text
(both divided by $\Delta \omega$) as function of the maximal
natural frequency $\Delta \omega$, in case of a one-dimensional
lattice with $N=100$ oscillators and for values of the symmetry
breaking parameter $\gamma = 0$ and $\gamma =2$. The raise in the
variance from zero indicates the transition to the desynchronized
phase.} \label{fig.former11}
\end{figure}

A further manifestation of the difference in the partial
synchronization patterns is seen in the average frequency $f$ and
the variance $\sigma$ as function of $\Delta\omega$ (Fig.
\ref{fig.former11}). Above the synchronization transition the
average frequency remains constant for $\gamma=0$, but increases
for $\gamma=2$ in agreement with Fig.s \ref{fig.former3} and
\ref{biforc2}, the variance increases in both cases ($\gamma=0$
and $\gamma=2$) above the transition, so that it may serve as
order parameter. Therefore we can tune the synchronization
features via the slope of the gradient.

The main qualitative feature, however, is in common to both
systems with and without antisymmetric coupling: the oscillator
with the highest natural frequency becomes the center of outgoing
circular waves, it is dynamically established as pacemaker. The
patterns, seen here in case of full and partial synchronization,
are quite similar to those predicted by Kuramoto \cite{kurabook}
for reaction-diffusion systems with continuous diffusion terms,
and to those experimentally observed in chemical oscillatory
systems \cite{zaikin}.

\section{Summary and conclusions}\label{sec.IV}
For pacemakers implemented as defect we have extended former
results on the entrainment frequency and the entrainment window to
arbitrary interaction range, analytically for $\alpha=0$,
numerically for intermediate $0<\alpha<\infty$. For large
dimensions the entrainment window decays exponentially with the
average distance of nodes from the pacemaker so that only shallow
networks allow entrainment. The synchronization transition is
reentrant as function of $\alpha$ (or $k$, the number of neighbors
on a ring, or $p$, the probability to add a random shortcut to the
ring topology). The entrainment gets most difficult for
$\alpha\sim 3$ and large $N$, while its most easily achieved for
next-neighbor and all-to-all couplings. This reentrance is easily
explained in terms of the normalization of the coupling strength.
For the same system without a pacemaker, but with a gradient in
the initial natural frequency distribution the oscillator with the
highest natural frequency becomes the center of circular waves,
its role as a pacemaker is dynamically induced without the need
for an asymmetric term in the interaction. Here we analytically
determined the synchronization transition as function of the
gradient slope. Above some threshold, full synchronization on one
or two-dimensional lattices is lost and replaced by partial
synchronization patterns. These patterns depend on the asymmetry
parameter $\gamma$. For $\gamma=0$ we observe a bifurcation in
frequency space, for $\gamma=2$ the one synchronized cluster
shrinks in its size, before for even steeper slopes
synchronization is completely lost. For artificial networks these
results may be used to optimize the placement and the number of
pacemakers if full synchronization is needed, or to control
synchronization by tuning the slope of natural frequency
gradients.

\end{document}